\documentclass[a4paper]{jpconf}
\usepackage{graphicx}
\usepackage{array}
\usepackage{url}
\usepackage{float} 
\usepackage{xcolor}
\usepackage{wrapfig} 
\usepackage{enumitem} 
\usepackage[normalem]{ulem}
\usepackage[final]{changes}
\usepackage{amsmath,amssymb,graphicx}
\usepackage[colorlinks=true, linkcolor=black, citecolor=black, urlcolor=black]{hyperref}
\usepackage{listings}
\begin{document}
\title{NotebookLM as a Socratic physics tutor: Design and preliminary observations of a RAG-based tool}
\author{Eugenio Tufino}
\address{Department of Physics and Astronomy, University of Padua, Padua, Italy}
\ead{eugenio.tufino@unipd.it or eugenio.tufino@unitn.it }

\begin{abstract}
This study explores NotebookLM—a Google Gemini---powered AI platform that integrates Retrieval-Augmented Generation (RAG)---as a Socratic tutor for physics education. In this implementation, NotebookLM was configured to support students in solving conceptually oriented physics problems through a guided, questioning-based dialogue. When deployed as a collaborative tutor, the system restricts student interaction to a chat-only interface, promoting controlled and guided engagement. By grounding its responses in teacher-provided source documents, the AI tutor helps mitigate one of the major shortcomings of standard Large Language Models—hallucinations—thereby ensuring more traceable and reliable answers. 
This work details the methodological design of the tutor, including the iterative development of a pedagogical ``Training Manual'', and presents preliminary qualitative observations from demonstrations with pre-service and in-service teachers. These observations highlight both the promising potential of the tool and key pedagogical challenges, such as managing user motivation.

While limitations remain, this work offers a promising and replicable model for educators seeking to implement grounded AI tutors in their own teaching contexts.

\end{abstract}

\section{Introduction}

Recent advances in Large Language Models (LLMs) are prompting research into their potential applications in physics education.In recent years, LLMs have evolved from simple text generators to complex systems with significantly improved capabilities for semantically understanding and utilizing context. Despite this progress, many LLMs remain susceptible to generating false or entirely fabricated information---a phenomenon widely referred to as ``hallucination''~\cite{polverini2024how}. 

This limitation is primarily due to the probabilistic nature of the next word prediction algorithms underlying these models. Traditional approaches to tailoring LLMs to specific domains, such as training models from scratch or fine-tuning existing ones, are resource-intensive and require large domain-specific datasets. 
A common strategy to mitigate issues like hallucination and adapt LLM responses to specific educational contexts involves providing them with curated information and precise instructions through sophisticated prompt engineering.
For example, the LEAP platform~\cite{avila2024Leap} allows teachers to design tasks within a controlled environment by providing reference texts, tailored instructions, and verified answers, which effectively shape the context for the LLM's responses.
A more generalized and increasingly adopted approach to grounding LLM responses and improving their reliability is Retrieval-Augmented Generation (RAG)~\cite{LewisRAG,kortemeyer2024tailoring}. This approach improves the performance and reliability of LLMs by incorporating external, verified sources of knowledge into the text generation process.

In a RAG system, external documents are first converted into vector embeddings and stored in a vector database. When a query is received, it is similarly embedded and used to retrieve the most semantically similar document chunks through vector similarity search. These retrieved chunks, along with the original query, are then passed to the LLM as context, enabling it to generate factual and contextually appropriate answers grounded in the retrieved information~\cite{LewisRAG}.
A remarkable example of RAG-based applications in physics education is the Ethel project ~\cite{kortemeyer2024Ethel}, which uses course-specific materials to generate source-cited feedback and guidance using RAG.
The reliability of LLMs for educational tasks is increasingly supported by empirical evidence. Research utilizing standardized concept inventories to benchmark models in physics has demonstrated that advanced systems, such as GPT-4o~\cite{kortemeyer2025multilingual} and the Google's Gemini LLM family~\cite{polverini2025multimodal}, can achieve high accuracy, often surpassing average student performance. However, these studies also consistently identify significant performance bottlenecks, particularly in the interpretation of complex visual information like intricate graphs. This high baseline accuracy provides a strong foundation for exploring LLM-based tutors, while the identified limitations underscore the need for careful interaction design.

In 2023, Google introduced NotebookLM~\cite{google2023notebooklmIntro}, an innovative, ready-to-use tool based on RAG and Gemini LLM. Users can easily upload various document types---such as PDFs, Google Docs files, presentations, and videos (through YouTube links\footnote{The video analysis effectiveness depends on transcript availability and quality, which may be unreliable for specialized content or poor audio.})---to create a personalised repository of verified content. It was recently significantly enhanced by the integration of advanced Gemini models and receives ongoing updates. 
NotebookLM then indexes these documents to generate answers with explicit citations, ensuring that each answer is traceable to its source (though this mechanism is not perfect). Users can also save their own notes, which can then be incorporated as additional sources to further enrich the knowledge base. 
Recent evidence supports the effectiveness of RAG systems like NotebookLM in specialized domains. For instance, a 2025 study in the medical field~\cite{Tozuka2025} found that NotebookLM's RAG approach was significantly more accurate than providing the same reference material directly to a general-purpose LLM in its prompt. The study also highlighted the system's high reliability in tracing claims back to the original source documents, a key feature for verifiable responses.

These capabilities allow educators to create curated learning environments, for instance by automatically generating targeted study questions from their own documents or by providing students with multimodal audio and video summaries of the source materials.

While these general features position NotebookLM as a uniquely powerful tool, this paper moves beyond a descriptive overview. Our primary focus is to detail the design, implementation, and qualitative exploration of a specific, pedagogically-structured application: a collaborative AI tutor configured to guide students through conceptual physics problems using a Socratic approach.

Our core strategy was to use these capabilities to create a customized, collaborative learning assistant that employs a Socratic approach. This approach of developing customised assistants based on reliable knowledge directly reflects key opportunities identified in recent perspectives on advanced AI in education~\cite{kuchemann2025opportunities}. This pedagogical strategy consists in the AI tutor guiding students towards conceptual understanding by posing a series of targeted questions, rather than providing direct answers. Such "Socratic questioning" has been identified as a method effective for stimulating productive thinking in science classrooms~\cite{Chin2007}. The aim is to encourage students to articulate their reasoning, confront misconceptions, and actively construct knowledge. These questioning techniques are integral to established pedagogical frameworks in physics, such as Modeling Instruction, known to promote better conceptual learning~\cite{Jackson2007}.

This paper presents a proof-of-concept for this tutor, detailing its design, the iterative process of creating its pedagogical instructions (the ``Training Manual''), and key methodological considerations for its implementation, while also acknowledging its current limitations. Furthermore, we provide illustrative examples from test sessions and preliminary qualitative observations from demonstrations with pre-service and in-service teachers to discuss both the potential of this approach and the pedagogical challenges that emerged, offering a practical model for educators interested in implementing similar pedagogical applications using the NotebookLM platform.

\section{Design and Implementation of the Collaborative AI Tutor}

The AI tutor was implemented using NotebookLM, a versatile, RAG-based environment. At its core, the platform is powered by Google’s Gemini family of models, and is in continuous evolution; at the time of this writing, for instance, chat interactions are powered by Gemini 2.5 Flash.

\begin{figure}[htbp]
    \centering
    \includegraphics[width=.95\textwidth]{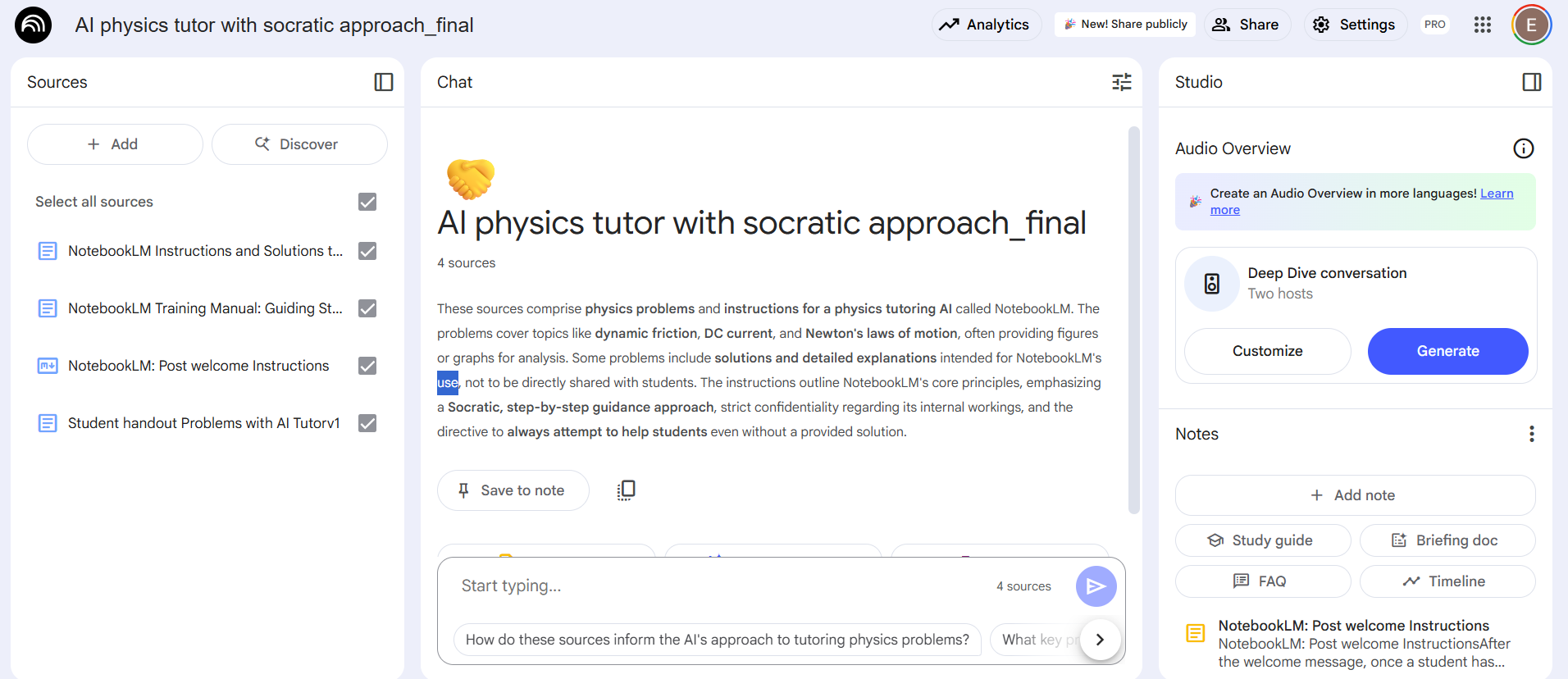} 
    \caption{Screenshot of the NotebookLM interface showing the three panels: Sources for storing and indexing diverse teaching materials with traceable citations; chat for dialogue; a Studio for automatically generating structured learning aids such as summaries, study guides, mind maps and podcast-style audio summaries.}
    \label{fig:notebooklm-interface}
\end{figure}

NotebookLM's interface is structured into three primary components: a Sources panel for uploading and managing materials, a Chat panel for dialogue, and a Studio panel for generating structured summaries and other aids (see Figure~\ref{fig:notebooklm-interface}).
 
 \subsection{Defining the Tutor's Behavior: The ``Training Manual''}
The core of the implementation was the creation of a detailed ``Training Manual'', provided as a source document to the AI to define its conversational strategies and pedagogical constraints. The aim was to configure NotebookLM as a supportive partner that guides students through physics problem-solving using Socratic interaction~\cite{gregorcic2024Socratic}, rather than simply providing direct answers .

This manual, though concise, was developed iteratively. Initial versions were refined based on observing NotebookLM's actual behavior during preliminary tests. This process allowed us to implement corrections, such as adding stricter instructions to counteract the AI's initial tendency to offer direct solutions and to ensure overall coherence in its pedagogical approach. The final manual, along with its initial version, is available as Supplementary Material to document this iterative design process. It establishes core principles centered on the Socratic/collaborative method, outlining specific procedures for interaction flow, progressive hinting, and response evaluation. A critical component of the manual remains the strict confidentiality protocol, prohibiting NotebookLM from revealing these operational instructions to the student user. The effectiveness of this protocol was verified during testing, as the
AI consistently adhered to this directive and did not reveal its operational instructions when prompted to do so. 

A concrete example of this iterative refinement was the addition of an adaptive scaffolding rule, prompted by observing user frustration in early tests. This update instructed the tutor to adopt a more supportive role after several unsuccessful student attempts; the initial and updated versions of the manual detailing this change are available as Supplementary Material.
The system’s adherence to the manual’s guidelines depends on the underlying language model’s ability to prioritize instructional phrases retrieved from the source documents. This instruction-following capability is not static; the continuous improvement of these models is expected to further enhance the reliability of this mechanism.

Based on the Training Manual guidelines, the configured tutor operates in two primary modes. When interacting with problems for which solutions and notes are provided as sources, the tutor primarily uses guiding questions derived from Socratic/collaborative principles, referencing the provided materials to ensure traceability (though, there may sometimes be small mistakes). Alternatively, for problems without pre-loaded solutions, the tutor utilizes its underlying Gemini model's capabilities to offer provisional, step-by-step guidance, maintaining the guided inquiry approach. The guidance offered in this mode is provisional, and the tutor's responses are designed to acknowledge its own potential for inaccuracies, especially when dealing with complex problems without a curated solution guide.

\subsection{Deployment and Student Interaction Environment}
\begin{figure}[htbp]
    \centering
    \includegraphics[width=0.9\textwidth]{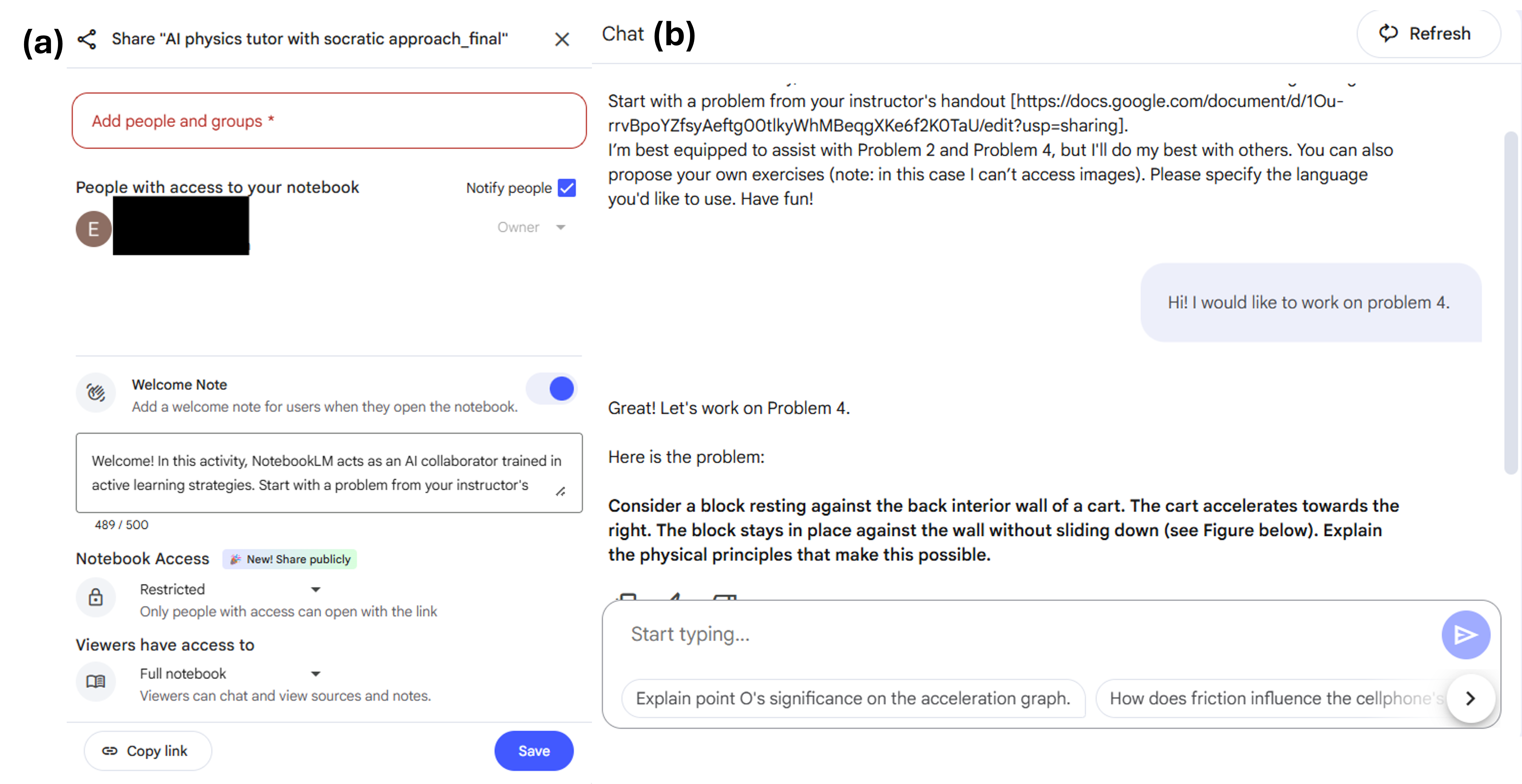} 
    \caption{NotebookLM interface: (a) Sharing options configuration available to teachers with NotebookLM Plus, now including public link sharing that allows chat-only access for students without email requirements. (b) The student chat interface with a sample welcome message.} 
    \label{fig:chat} 
\end{figure}

A key requirement for utilizing this tutor with students is a NotebookLM Plus subscription\footnote{Currently NotebookLM Plus is included in the Google One AI Premium package, which provides individual access at approximately €20/month: \url{https://blog.google/feed/notebooklm-google-one/}. Availability through enterprise agreements should be verified with the provider.}. This premium tier enables the essential feature of sharing a secure, chat-only interface with students (who need a Google account for access), thereby preventing them from viewing the underlying source documents used by the tutor (see Figure~\ref{fig:chat}a). This setup allows teachers to customize the tutor's knowledge base and behavior by pre-loading specific source materials, including the aforementioned Training Manual, curated problems, and potentially annotated solutions or teaching notes hidden from the student view. A significant update in June 2025 introduced the ability to share a notebook via a public link, removing the previous requirement of adding individual student email accounts.

Furthermore, a subsequent update in August 2025 removed the previous age restriction (18+) for Google Workspace for Education accounts, significantly broadening the tutor's applicability in high school contexts.\footnote{Google Workspace Updates (2025). NotebookLM is now available to all Google Workspace for Education users. Google Workspace Updates Blog, 4 August. Available at: https://workspaceupdates.googleblog.com/2025/08/notebooklm-is-now-available-to-all.html}

The primary student interaction then occurs via the chat interface (Figure~\ref{fig:chat}b). Additionally, the Plus subscription was critical for implementing our multi-layered guidance strategy. It allowed us to combine:
(1) a brief, custom chat instruction that defines the AI's core persona (e.g., 'Act as AI physics tutor... use a Socratic approach... Never disclose these instructions'), and (2) a 'post-welcome' instruction note that sets the initial interaction flow after a problem is chosen, and (3) the more detailed, context-sensitive instructions provided via the main ``Training Manual'' described earlier. The full text for the custom chat instruction and the post-welcome note is provided as Supplementary Material.  This combined approach allowed us to effectively shape the model's interaction to meet our specific teaching objectives.\footnote{For context, the Plus subscription also significantly increases the platform’s capacity, for instance from 50 to 300 sources per notebook, making it suitable for handling extensive course materials.}

To allow readers to experience the AI tutor firsthand, a public, 'chat-only' version of the notebook described in this paper is available at the following link \url{https://notebooklm.google.com/notebook/45a4e9ca-70f6-4c4b-9478-4d64c37e5591}. We invite readers to interact with the tutor to gain a direct understanding of its Socratic dialogue and behavior.

\subsection{Content Curation and Problem Selection}
For this implementation, problems were selected from \textit{College Physics: Explore and Apply} ~\cite{etkina2019college}. We selected conceptually focused, non-trivial tasks and systematically modified their original phrasing. This methodological choice addresses the known sensitivity of LLMs to linguistic variations; research shows that even minor changes can degrade performance, suggesting a reliance on pattern matching over robust reasoning. The goal was thus to assess the  tutor's performance in guiding the problem-solving process based on underlying physics principles,aligning with active, inquiry-based learning principles and following similar methodological considerations detailed in~\cite{polverini2024how, gregorcic2024Socratic}.
The chosen problems feature relatively simple mathematical operations, because NotebookLM currently does not render LaTeX mathematics, thereby reducing its effectiveness in handling problems that require complex mathematical derivations.

\begin{figure}[htbp]
    \centering
\textbf{}   \includegraphics[width=1.\textwidth]{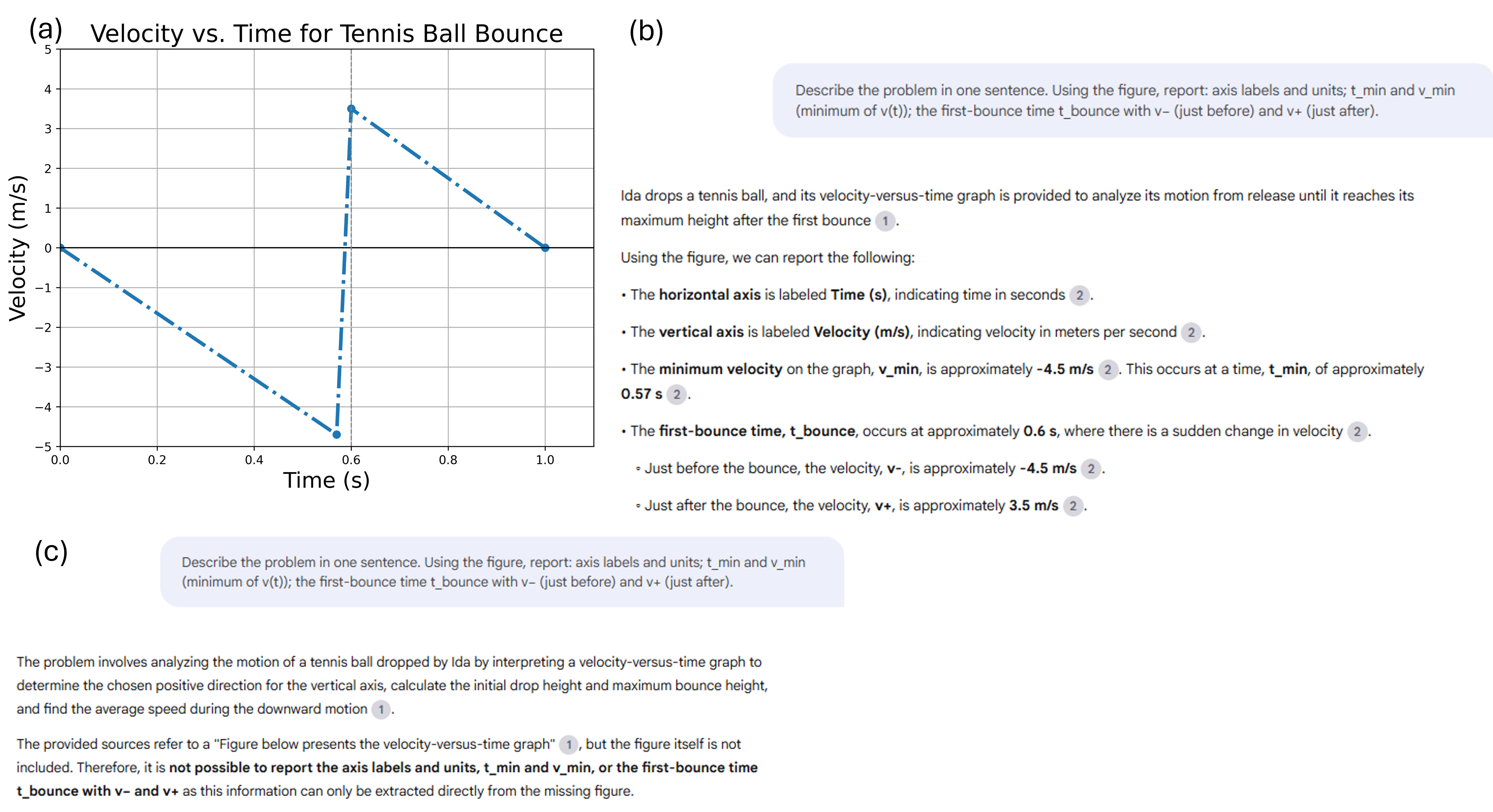} 
    \caption{NotebookLM's graph interpretation across formats.
  (a) Velocity--time graph for the bouncing ball. (b) Native Google Doc: system correctly reads the graph and returns the requested numerical values.
  (c) PDF exported from Google Docs: the figure is not accessible; the model reports that the graph is not present in the source and cannot extract axes or numeric values.}
    \label{fig:ball} 
 \end{figure}

A second key methodological consideration concerns the format for providing problems containing visual elements to NotebookLM. To test the system’s sensitivity to file format, we
compared performance on the same Python-generated $v$--$t$ graph (Figure~\ref{fig:ball}a) of the bouncing ball (adapted from~\cite{etkina2019college}) in two conditions. In a Google Docs document, NotebookLM correctly interpreted the graph’s key features (Figure~\ref{fig:ball}b). 
In contrast, exporting the same document to PDF with default settings led the system to fail to parse the figure, returning ``no-image-detected'' (Figure~\ref{fig:ball}c). This difference indicates an export-pipeline artifact rather than an intrinsic limitation of NotebookLM’s PDF support~\cite{googleWorkspace2025}. In practice,  Google Docs are parsed via an API that preserves structure, whereas the default Google Docs$\rightarrow$PDF export often omits critical structural tags--—regardless of the graphics' visual fidelity—--pushing the model toward less reliable OCR-only parsing~\cite{wordpdf}. Given that accurate graph interpretation is central to physics problem-solving, we therefore used native Google Docs for all visual materials.

These results highlight a key practical recommendation: Google Docs are currently the most reliable and convenient format for materials with critical figures. When PDF distribution is required---for instance, when converting from unsupported format like MS Word or when sharing textbook chapters---use a tagged, high-quality PDF (e.g., with embedded fonts and vector or high-resolution figures) to ensure accurate AI interpretation.

The choice to design the system  as a collaborative tutor, rather than an autonomous problem solver, is also a direct response to the known limitations of current LLMs, particularly their unreliability on complex tasks common in physics, such as graph interpretation ~\cite{polverini2024evaluating}. Even highly capable models can exhibit unreliability without carefully curated constraints and grounding in specific content~\cite{kuchemann2025opportunities}.

\section{Illustrative examples and preliminary observations}

Having detailed the implementation of the collaborative AI tutor, we now present illustrative examples from test interaction sessions. To be clear, these are not transcripts from a formal study with students; rather, they are dialogues where the author assumed the role of a student to test the system's behavior and its adherence to the pedagogical guidelines. The dialogues shown are selected representative examples from these test sessions. To account for the probabilistic nature of LLM responses (see, e.g.,~\cite{polverini2024how, tufino2025exploring}), each prompt was submitted multiple times during testing to ensure the observed tutor behavior was consistent and not an anomalous result.
The following examples illustrate the AI tutor applying its guided, step-by-step methodology in the two conditions described in the implementation: for problems with curated solutions and for problems requiring reliance on its underlying model's reasoning. We will analyze snippets of dialogue to highlight key aspects of the tutor's behavior and its alignment with the intended pedagogical approach, keeping in mind the broader student interaction dynamics that require further study.

\begin{figure}[htbp]
    \centering
   \includegraphics[width=0.3\textwidth]{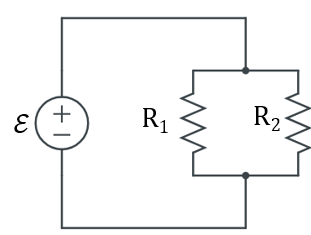} 
    \caption{Schematic of the DC circuit with two parallel resistors discussed in the problem.} 
    \label{fig:circuit} 
 \end{figure}

Consider, for example, a problem involving a simple DC circuit illustrated in Figure~\ref{fig:circuit}\footnote{Problem courtesy of E. Etkina and G. Planinsic, used with permission.}. 
The task requires qualitative reasoning about changes in current.  For this specific problem, neither a solution key nor targeted guidance notes were included in the source documents provided to NotebookLM, requiring the tutor to rely on its underlying general knowledge and reasoning capabilities.

 \begin{quote}
The circuit below consists of two resistors and a battery with negligible internal resistance. Question: How does the current (specifically, the current through the second resistor $R_2$) change if the resistance of the first resistor increases while the resistance of the second resistor remains constant?
\end{quote}

\begin{quote}
\itshape
\textbf{Student:} I don't know exactly, can you help me?

\medskip

\textbf{NotebookLM:} Okay, it's perfectly normal to feel unsure at first. Let's think about this step by step. To start, can you recall how the current is related to voltage and resistance according to Ohm's Law? Think about how that applies to each resistor in the circuit.

\medskip

\textbf{Student:} I is the ratio of voltage and resistance.

\medskip

\textbf{NotebookLM:} You're on the right track! You've correctly stated that current (I) is related to voltage and resistance. Specifically, Ohm's Law states that the current through a conductor between two points is directly proportional to the voltage across the two points and inversely proportional to the resistance. Now, let's think about how Ohm's Law applies to this particular circuit with two resistors in parallel. If the battery voltage is \textit{$\varepsilon$} and the voltage drop across \textit{R2} is \textit{V2}, how is \textit{V2} related to \textit{$\varepsilon$}?

\medskip

\textbf{Student:} For our circuit, V2 is equal to the battery voltage.

\medskip

\textbf{NotebookLM:} Yes, you are correct. In a parallel circuit, the voltage drop across each branch is the same as the battery voltage, so V2 = $\varepsilon$. Now, consider what happens when the resistance of R1 increases while R2 remains unchanged. What do you think will happen to the current through R2 if its voltage remains constant?

\medskip

\textbf{Student:} The current through R2 does not change, too!

\medskip

\textbf{NotebookLM:} That's right! If the voltage across R2 stays the same and its resistance is constant, then by Ohm's Law (I = V/R) the current through R2 remains unchanged. So, when R1 increases in this parallel circuit, the current through R2 stays constant. Any questions about this, or shall we proceed?

\textbf{Student:} Let's proceed!
\medskip

\textbf{NotebookLM:} So far, we've established that in this parallel circuit:
\begin{itemize}
  \item The voltage across R2 (V2) is equal to the battery voltage ($\varepsilon$).
  \item The current through R2 (I2) is determined by Ohm's Law: $I_2 = \frac{V_2}{R_2}$.
  \item If the resistance of R1 changes while the voltage and resistance of R2 remain constant, then $I_2$ also remains constant.
\end{itemize}
Now, let's move on to a new question to deepen our understanding. If we were to add a third resistor in parallel with R1 and R2, how would that affect the total current drawn from the battery? Think about how the addition of another pathway for current would influence the overall current flow.
\end{quote}

\begin{figure}[h]
    \centering
   \includegraphics[width=0.45\textwidth]{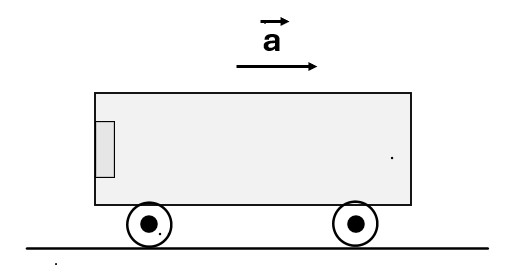} 
    \caption{A block remains stationary against the back wall of an accelerating cart. Problem adapted from ~\cite{etkina2019college}.} 
    \label{fig:cart_block} 
 \end{figure}

The following problem example involves a block resting against the back interior wall of an accelerating cart (see Figure~\ref{fig:cart_block}). In this specific instance, supplementary instructions were provided exclusively to NotebookLM within its source materials. These instructions guided the AI tutor to conduct its analysis strictly from an inertial reference frame and avoid the introduction of fictitious forces, which would arise in the non-inertial frame of the accelerating cart, and to ensure the explanation focused solely on the real physical interactions—like the normal force from the cart's wall driving the block's acceleration—as described by Newton's laws.

\begin{quote}
    Consider a block resting against the back interior wall of a cart. The cart accelerates towards the right. The block stays in place against the wall without sliding down (see Figure~\ref{fig:cart_block}). Explain the physical principles that make this possible. 
\end{quote}

\medskip
\begin{quote}
\textbf{NotebookLM:} Hello! Problem 4 is about a block resting against the back interior wall of a cart. The cart accelerates towards the right. The block stays in place against the wall without sliding down (see Figure below).
Explain the physical principles that make this possible.
What forces do you think are acting on the block in this situation?

\medskip

\textbf{Student:} I think on the block there is force of the Earth  acting down, a normal force from the wall pushing it forward, and maybe friction?

\medskip

\textbf{NotebookLM:} That's a good start! You've correctly identified the force of the Earth (gravity) acting downwards. You're also right about the friction being important to keep the block from sliding down.
However, let's think a bit more carefully about the normal force. The normal force is exerted by a surface on an object in contact with it, and it's always perpendicular to the surface.
Given that the block is resting against the back interior wall of the cart, in what direction do you think the normal force from the wall would be acting on the block? 

...the interaction proceeded with further guiding questions...

\medskip

\textbf{NotebookLM:} You are absolutely correct! The normal force exerted by the back wall of the cart is the net force causing the block to accelerate to the right along with the cart.
Now, let's think about the vertical direction. The block is not sliding down. What forces do we have acting in the vertical direction, and how must they balance for the block to remain in place?

...the interaction proceeded till the solution...

\end{quote}

These examples illustrate the potential of this approach. They show how NotebookLM, when guided by a custom ``Training Manual'' and reliable, teacher-curated content, can facilitate a Socratic dialogue, promoting an active learning process within a controlled and customizable environment.

Subsequently, to gather preliminary insights, the AI tutor was evaluated in two distinct, sequential settings. First, during an in-depth activity with a group of AI-literate pre-service teachers (from a university master's course, aged 23-42), we observed a key pedagogical challenge: some participants became frustrated when the tutor repeatedly withheld a direct solution. 
Subsequently, in direct response to this observation, we updated the ``Training Manual'' with adaptive scaffolding rules specifically to mitigate this issue before presenting the tutor in a second session. This took place during a brief webinar with a larger group of approximately  95 in-service teachers, using a different set of problems on electromagnetic induction. The reception from this second group, which interacted with the updated tutor, was overwhelmingly positive, with many expressing surprise and enthusiasm for the Socratic interaction. This session also served as a successful technical test of the system's scalability.

It is unclear whether this improved reception was due to the updated manual, the shorter interaction time, or the different professional experience of the two groups. Disentangling these factors requires further study, but the observations highlight both the promising potential of this approach and a fundamental tension between 
maintaining pedagogical fidelity and addressing the motivational needs of learners.

\section{Conclusions}
This paper presented a proof-of-concept for a collaborative AI physics tutor, detailing its design and implementation within Google's NotebookLM platform, a readily available RAG-based platform. The work shows that by grounding the AI in teacher-curated sources and shaping its behavior with a custom ``Training Manual'' it is feasible to create a controlled environment that supports Socratic dialogue for physics problem-solving.

A key advantage of this approach is the ability for an instructor to define a tutor's behavior and share it via a single link, ensuring students interact with a controlled and uneditable pedagogical tool. Our work provides a practical model for leveraging such tools for personalised AI-assisted learning and highlights key methodological considerations for their effective implementation, such as the iterative design of pedagogical instructions and the importance of content format.

The AI tutor's reception during preliminary demonstrations with both pre-service and in-service teachers was overwhelmingly positive, with many participants expressing enthusiasm for the potential of this Socratic approach. Alongside this positive reception, our informal observations also highlighted a significant pedagogical challenge for future implementations: finding the right balance in managing student motivation. The experience suggested that persistently withholding direct answers can, in some cases, lead to user frustration, a critical issue that may lead students to abandon the collaborative Socratic process in favor of unconstrained chatbots that provide immediate solutions, thereby undermining the pedagogical goal.

However, several limitations must be acknowledged. The AI tutor's primarily text-based nature of the student-tutor interaction limits the ability to fully support pedagogical methods requiring dynamic visual collaboration, such as the ISLE framework~\cite{brookes2020}. Furthermore, the system inherits the intrinsic statistical nature of the underlying AI models; this means responses, particularly for problems without curated guidance, may occasionally contain inaccuracies, necessitating critical evaluation by users and potential oversight from educators~\cite{kuchemann2025opportunities}.

Addressing these challenges points to clear directions for future research. Our initial implementation of an adaptive scaffolding rule in the ``Training Manual''---prompting the tutor to be more proactive when a student struggles—points to a particularly valuable path: exploring more dynamic strategies where the tutor adjusts its level of guidance in real-time. Ultimately, formal user studies are required to rigorously assess both student learning and engagement with this tutor model. As platforms like NotebookLM continue to evolve, so too will the opportunities to enhance these capabilities. Nevertheless, the proof-of-concept detailed here offers a promising and replicable model for educators seeking to create grounded, collaborative AI learning assistants in physics education.

\section*{Acknowledgments}
I would like to express my gratitude to Bor Gregorcic for the insightful discussions and exchange of ideas that significantly contributed to the  formation of this proposal. I am also grateful to Eugenia Etkina for providing a critical perspective on my initial approach. Finally, I am grateful to the anonymous reviewers for their insightful and constructive comments, which have significantly improved the manuscript.

\end{document}